\documentclass[a4paper,dvips,12pt]{article}
\usepackage{epsfig,graphics,graphicx}
\begin{document}

\pagestyle{myheadings}
\markright{\it P233}
\vskip.5in
\begin{center}

%
%
\vskip.4in {\Large\bf An alternative calculation of the Casimir energy for kappa-deformed electrodynamics}
\vskip.3in
%
%
%
D Pinheiro\footnote{Email: \tt daniel@if.ufrj.br }\\
F C Santos\footnote{Email: \tt filadelf@if.ufrj.br }\\
A C Tort\footnote{Email: \tt tort@if.ufrj.br}\\
Departamento de F\'{\i}sica Te\'{o}rica, Instituto de F\'{\i}sica,\\
Universidade Federal do Rio de Janeiro, CP 68528,\\
21945-970 Rio de Janeiro, Brasil.

%
%
%
\end{center}
%
\vskip.2in
\begin{abstract}
A simple, but effcient way of calculating regularized Casimir energies suitable for non-trivial frequency spectra is briefly described and applied to the case of a kappa-deformed scalar field theory. The results are consistent with the ones obtained by other means.

\end{abstract}
%
%
%
%
%

\section{Introduction}
The evaluation of Casimir energies \cite{Casimir} associated with confined zero-point fluctuations of quantum fields depends on several factors, the more relevant ones are the nature of the quantum field, the type of space-time manifold and its dimensionality, and the type of boundary conditions imposed on the fields. All these factors taken together can lead to relatively simple dispersion relations, e.g.: a massless scalar field under Dirichlet b.c., or to rather complex ones, e.g.: a massive fermion field under MIT b.c.. Recently a very simple,  but rather powerful way of calculating regularized Casimir energies suitable for non-trivial frequency spectra was introduced in the literature \cite{IJMPA2003}. This technique relies on well-known theorems of complex analysis, to wit: the Cauchy integral formula and the Mittag-Leffler expansion theorem in one of its simplest versions. The method was successfully employed in the evaluation of the Casimir energy of a massive fermionic field confined in a $d+1$-dimensional slab-bag \cite{IJMPA2003} and also in the evaluation of the Casimir energy of confined bosonic and fermionic fields in the presence of pertinent b.c. and a uniform magnetic field \cite{JPA2003}. Here we will sketch the application of this method to the evaluation of another example of a sum over a non-trivial frequency spectrum, namely: the one associated with the regularized Casimir energy of a kappa-deformed scalar field.
\section{A sum formula}
Consider a quantum field in a $d+1$-dimensional flat space-time under b.c. imposed on two hyperplanes of area $L^{d-1}$. The distance between the hyperplanes is $\ell$, hence there is a restriction on one of the spatial dimensions, the $OX_d$ axis. Suppose $L\gg \ell$. At the one-loop level, the (unregularized) Casimir energy is given by
\begin{equation}\label{}
    E_0\left(d\right)=\alpha\left(d\right)\frac{L^{d-1}}{2}\int\sum_n
    \frac{d^{d-1}}{\left(2\pi\right)^{d-1}}\Omega_n,
\end{equation}
where $\alpha\left(d\right)$ is a dimensionless factor that takes into account the internal degrees of freedom of the field and
\begin{equation}\label{}
    \Omega_n=\sqrt{p_\bot^2+\frac{\lambda^2_n}{\ell^2}+m^2},
\end{equation}
where $p_\bot^2=p_1^2+p_2^2+\cdots +p_{d-1}^2$, $m$ is the mass of an elementary excitation of the field and $\lambda_n$ is the $n$th real root of the transcendental equation determined by the b. c.. It is shown in Ref. \cite{IJMPA2003} that Cauchy integral formula and Mittag-Leffler expansion theorem carefully applied allow us to recast  the sum above into the form
\begin{equation}\label{}
E_0\left(d\right)=\alpha\left(d\right)\frac{L^{d-1}}{2}\int\frac{d^d\,p}
{\left(2\pi\right)^d}\log\left[1+\frac{K_1\left(z\right)}
{K_2\left(z\right)}\right],
\end{equation}
where $z$ is defined by
\begin{equation}\label{}
    q^2+\Omega_n^2=\frac{z^2+\lambda_n^2}{\ell^2},
\end{equation}
with $q$ as an auxiliary complex momentum. The functions $K_1\left(z\right)$ and $K_2\left(z\right)$ must be constructed from the boundary conditions. This last relation gives the regularized Casimir energy. For details see \cite{IJMPA2003}, and see also \cite{JPA2003} for a simple example. It can be shown \cite{Santos&Tort2004} that this method can be extended to the case where the sum we must evaluate is performed over a function of $\Omega_n$, i.e.:
\begin{equation}\label{}
    E_0\left(d\right)=\alpha\left(d\right)\frac{L^{d-1}}{2}\int\sum_n
    \frac{d^{d-1}}{\left(2\pi\right)^{d-1}}f\left(\Omega_n\right).
\end{equation}
This is precisely what happens when we consider a kappa-deformed scalar theory. Kappa-deformed theories are theories whose associated symmetry groups depart from the usual symmetry groups of standard quantum field theories, e.g.: the Poincar\'e group. This departure is measured by some convenient parameter, say $\kappa$.
\section{The Casimir effect for kappa-deformed scalar field theory}

The kappa-deformed Klein-Gordon equation in $3+1$ dimensions is given by (see \cite{tese} and references therein)
\begin{equation}\label{}
\left[\vec{\nabla}^2-\left(2\kappa\sinh\left(\frac{P_0}{2\kappa}\right)\right)^2
-m^2\right]\phi\left(\vec{x},x_0\right)=0,
\end{equation}
where $\kappa$ is a parameter with dimensions of a mass. The on shell condition reads
\begin{equation}\label{}
\vec{P}^{\,2}-\left(2\kappa\sinh\left(\frac{P_0}{2\kappa}\right)\right)^2=-m^2,
\end{equation}
Upon diagonalizing simultaneously the operators $\vec{P}$ and $P_0$ we obtain a new frequency spectrum which is given by
\begin{equation}\label{}
\omega\left(\vec{p}\right)=2\kappa\sinh^{-1}\left(\frac{1}{2\kappa}
\sqrt{\vec{p}^{\,2}+m^2}\right) \end{equation}
Suppose Dirichlet boundary conditions are imposed on two parallel surfaces perpendicular to the $OX_3$-axis. Then
\begin{equation}
E_0\left(\ell\right)=\frac{L^2}{\eta}\int\frac{d^2\,p_\bot}
{\left(2\pi\right)^2}\sum_n\sinh^{-1}\left(\eta\,\Omega_n\right),
\end{equation}
where for convenience we have defined $\eta:=1/2\kappa$. By applying the techniques described in \cite{IJMPA2003} to this more complex case we obtain
\begin{equation}\label{ten}
{E}_{0}{(\ell)=-}\frac{L^{2}}{\eta}\int \frac{d^{2}\,p_{\bot}}{%
(2\pi )^{3}}\oint {dq}{\sum}_n\frac{2q^{2}}{q^{2}+\Omega
_{n}^{2}}\frac{\sin ^{-1}(\eta q)}{q},
\end{equation}
where as before $q$ plays the role of an auxiliary momentum. The pole structure of the above equation is the same as in the absence of deformation, moreover, we can consider the boundary conditions as being the same. In our case this will mean Dirichlet boundary conditions applied to the undeformed case. 

It follows from Eq. (\ref{ten}) that
\begin{equation}\label{eleven}
E_{0}(\ell)=L^{2}\int\frac{d^{2}p_{\bot}}{(2\pi)^{3}}%
\int_{0}^{\frac{1}{\eta }}\frac{dq}{\sqrt{1-\eta ^{2}q^{2}}}\log
\left[1+\frac{K_{1}\left(z\right)}{K_{2}\left(z\right)}\right]
\end{equation}
where $K_1\left(z\right)$ and $K_2\left(z\right)$ are two functions that can be read out from the boundary conditions \cite{IJMPA2003}. For Dirichlet b.c. we have $K_{1}\left(z\right)=-e^{-z}/2$ and $K_{2}\left(z\right)=e^{z}/2$, with $z=\ell\sqrt{\vec{p}_\bot^{\,2}+q^2}$ \cite{JPA2003}. Taking these two functions into Eq. (\ref{eleven}) we obtain

\begin{equation}
E_0\left(\ell\right)=\frac{L^2}{8\pi^3}\int {d}^2
p_\bot\int_0^{\frac{1}{\eta }}\frac{dq}{\sqrt{1-\eta ^2q^2}}
\log\left[1-e^{-2\ell\sqrt{\vec{p}_\bot^{\,2}+q^2}}\right],
\end{equation}
which is in agreement with \cite{tese}. If we take the limit $\eta \to 0$ after some manipulations we obtain
\begin{equation}
E_0\left(\ell\right)=\frac{L^2}{2\pi^2}\ell^3\int_0^\infty
dz\,z^2\log\left[1-e^{-2z}\right],
\end{equation}
which leads to the original Casimir energy for this situation \cite{JPA2003}.

\section{Final remarks}
In this brief note we described an alternative method for summing over non-trivial frequency spectra and applied the main result to the problem of evaluating the Casimir energy of a kappa-deformed scalar theory under Dirichlet boundary conditions. The results obtained are consistent with the ones obtained by other methods and hold for quantum electrodynamics between parallel plates. This method can be also applied to other non-trivial examples. A more detailed account of the extension of the technique employed in the  example discussed here is in preparation and the results will be published elswhere. 
\section*{Acknowledgments}
D. Pinheiro acknowledges the financial support of the Instituto de F\'{\i}sica, UFRJ, graduate program; F. C. Santos and A. C. Tort acknowledge the financial support of XXIV ENFPC Organizing Comitee that allowed them to be part of the XXIV Brazilian National Meeting on Particles and Fields held in Caxambu MG, Brazil, September 30 -- October 4, 2003.

%
%
%
%
%

%
\end{document}